\documentstyle[aps]{revtex}

\def\bfg{\begin{figure}}
\def\efg{\end{figure}}

\def\be{\begin{equation}}
\def\ee{\end{equation}}

\begin{document}
\draft
\title{Collective motion of organisms in three dimensions
}
\author{Andr\'as Czir\'ok$^{1}$,
M\'aria Vicsek$^2$ and Tam\'as Vicsek$^{1}$
}

\address{
$^1$  Department of Atomic Physics,
E\"otv\"os University, Budapest, Puskin u. 5-7, 1088 Hungary \\
$^2$ Department of Computational Mathematics, School of Economics,
Budapest, CIMCIM, and
Computer and Automation Institute of HAS, Budapest, P.O.B.
63, 1518 Hungary
\\ }

\maketitle

\begin{abstract}

We study a model of flocking in order to describe the transitions during
the collective motion of organisms in three dimensions (e.g., birds).
In this model the particles representing the organisms are
self-propelled, i.e., they move with the same absolute velocity $v_0$.
In addition, the particles locally interact by choosing at each time
step the average direction of motion of their neighbors and the effects
of fluctuations are taken into account as well.  We present the first
results for large scale flocking in the presence of noise in three
dimensions.  We show that depending on the control parameters both
disordered and long-range ordered phases can be observed.  The
corresponding phase diagram has a number of features which are
qualitatively different from those typical for the analogous equilibrium
models.

 \end{abstract}

\section{Introduction}

The collective motion of organisms (birds, for example), is a
fascinating phenomenon many times capturing our eyes when we observe our
natural environment.  In addition to the aesthetic aspects of collective
motion, it has some applied aspects as well: a better understanding of
the swimming patterns of large schools of fish can be useful in the
context of large scale fishing strategies.  In this paper we address the
question whether there are some global, perhaps universal transitions in
this type of motion when many organisms are involved and such parameters
as the level of perturbations or the mean distance of the organisms is
changed.

Our interest is also motivated by the recent developments in areas
related to statistical physics.  During the last 15 years or so there
has been an increasing interest in the studies of far-from-equilibrium
systems typical in our natural and social environment.  Concepts
originated from the physics of phase transitions in equilibrium systems
\cite{HES71} such as collective behaviour, scale invariance and
renormalization have been shown to be useful in the understanding of
various non-equilibrium systems as well.  Simple algorithmic models have
been helpful in the extraction of the basic properties of various
far-from-equilibrium phenomena, like diffusion limited growth
\cite{DLA}, self-organized criticality \cite{SOC} or surface roughening
\cite{surface}.  Motion and related transport phenomena represent a
further characteristic aspect of non-equilibrium processes, including
traffic models \cite{traffic}, thermal ratchets \cite{MM} or driven
granular materials \cite{gran}.

Self-propulsion is an essential feature of most living systems.  In
addition, the motion of the organisms is usually controlled by
interactions with other organisms in their neighbourhood and randomness
plays an important role as well.  In Ref. \cite{VCBCS95} a simple model
of self propelled particles (SPP) was introduced capturing these
features with a view toward modelling the collective motion of large
groups of organisms \cite{Reynolds87} such as schools of fish, herds of
quadrupeds, flocks of birds, or groups of migrating bacteria
\cite{AH91,bacbio,film}, correlated motion of ants \cite{millonas} or
pedestrians \cite{Helbing}.  Our original SPP model represents a
statistical physics-like approach to collective biological motion
complementing models which take into account much more details of the
actual behaviour of the organism, but, as a consequence, treat only a
moderate number of organisms and concentrate less on the large scale
transitions \cite{Reynolds87,jpn}.

In this paper the large scale transitions during collective motion in
three dimensions is considered for the first time.  Interestingly,
biological motion is typical in both two and three dimensions, because
many organisms move on surfaces (ants, mammals, etc), but can fly
(insects, birds) or swim (fish).  In our previous publications we
demonstrated that, in spite of its analogies with the ferromagnetic
models, the transitions in our SSP systems are quite different from
those observed in equilibrium models.  In particular, in the case of
equilibrium systems possessing continuous rotational symmetry the
ordered phase is destroyed at finite temperatures in two dimensions
\cite{MW66}.  However, in the 2d version of the non-equilibrium SSP
model phase transitions can exist at finite noise levels (temperatures)
as was demonstrated by simulations \cite{VCBCS95} and by a theory based
on a continuum equation developed by Toner and Tu \cite{TT}.  Thus, the
question of how the ordered phase emerges due to the non-equilibrium
nature of the model is of considerable theoretical interest as well.

In section 2 we describe our model. The results are presented in section
3 and the conclusions are given in section 4.

\section{Model}

The model consists of particles moving in three dimensions with periodic
boundary conditions.  The particles are characterised by their
(off-lattice) location $\vec{x}_i$ and velocity $\vec{v}_i$ pointing in
the direction $\vartheta_i$.  To account for the self-propelled nature
of the particles the magnitude of the velocity is fixed to $v_0$.  A
simple local interaction is defined in the model: at each time step a
given particle assumes the average direction of motion of the particles
in its local neighbourhood $S(i)$ with some uncertainty, as described by

\begin{eqnarray}
\vec{v}_i(t+\Delta t)=\hbox{ \bf N}(\hbox{ \bf N}(\langle
\vec{v}(t)\rangle_{S(i)}) + \vec\xi),\\
\label{EOM3D}
\end{eqnarray}
where $\hbox{ \bf N}(\vec{u})=\vec{u}/\vert\vec{u}\vert$ and the noise
$\vec\xi$ is uniformly distributed in a sphere of radius $\eta$.

The positions of the particles are updated according to
\begin{equation}
\vec{x}_i(t+\Delta t) = \vec{x}_i(t) + v_0\vec{v}_i(t)\Delta t,
\label{update}
\end{equation}

The model defined by Eqs. (\ref{EOM3D}) and (\ref{update}) is a
transport related, non-equilibrium analogue of the {
ferromagnetic} models.
The analogy is as follows: the Hamiltonian tending to align the spins in the
same direction in the case of equilibrium ferromagnets is replaced by
the rule of aligning the direction of motion of particles, and  the
amplitude of the random perturbations can be considered proportional to
the temperature.

We studied this model
 by performing  Monte-Carlo simulations.  Due to the
simplicity of the model, only two control parameters should be
distinguished: the (average) density of particles $\varrho$ and the
amplitude of the noise $\eta$.  In the simulations random initial
conditions and periodic boundary conditions were applied.

\section{Results}

For the statistical characterisation of the configurations, a
well-suited order parameter is the magnitude of the average momentum of
the system: $\phi\equiv\left\vert \sum_j \vec{v}_j \right\vert/N$.  This
measure of the net flow is non-zero in the ordered phase, and vanishes
(for an infinite system) in the disordered phase.

The simulations were started from a disordered configuration, thus
$\phi(t=0)\approx 0$.  After some relaxation time a steady state emerges
indicated, e.g., by the convergence of the cumulative average $(1/\tau)
\int^\tau_0 \phi(t)dt$.

The stationary values of $\phi$ are plotted in Fig.~1. vs $\eta$ for
$\varrho = 2$ and various system sizes $L$ (indicated in the plot by the
number of particles).  For weak noise the model displays long-range
ordered motion (up to the actual system size $L$) disappearing in a
continuous manner by increasing $\eta$.

These numerical results suggest the existence of a kinetic phase
transition as $L \rightarrow \infty$ described by

\begin{equation}
\phi(\eta)\sim \cases{
         \Bigl({\eta_c(\varrho) - \eta\over \eta_c(\varrho)}\Bigr)^\beta
                & for $\eta<\eta_c(\varrho)$ \cr
        0  & for $\eta>\eta_c(\varrho)$ \cr
    },
\label{scale}
\end{equation}

where $\eta_c(\varrho)$ is the critical noise amplitude that separates
the ordered and disordered phases. Due to the nature of our
non-equilibrium model it is difficult to carry out simulations on a
scale large enough to allow the precise determination of the critical
exponent
$\beta$. We find that the exponent 1/2 (corresponding to the mean
field result for equilibrium magnetic systems) fits our results within
the errors. This fit is shown as a solid line.

Next we discuss the role of density.  In Fig.~2a, $\phi(\eta)$ is
plotted for various values of $\varrho$ (by keeping $N$=Const. and
changing $L$).  One can observe that the long-range ordered phase is
present for any $\varrho$, but for a fixed value of $\eta$, $\phi$
vanishes with decreasing $\varrho$.  To demonstrate how much this
behaviour is different from that of the diluted ferromagnets we have also
determined $\phi(\eta)$ for $v_0=0$.  In this limit our model reduces to
an equilibrium system of randomly distributed "spins" with a
ferromagnetic-like interaction.  This system is analogous to the three
dimensional diluted Heisenberg model.  In Fig.~2b we display the results
of the corresponding simulations.  There is a major difference between
the self-propelled and the static models: in the static case the system
{\it does not order} for densities below a critical value close to 1
which in the units we are using corresponds to the percolation threshold
of randomly distributed spheres in 3d.

This situation is demonstrated in the phase diagram shown in Fig~3. Here
the diamonds show our estimates for the critical noise for a given
density for the SPP model and the crosses show the same for the static
case. The SPP system becomes ordered in the whole region below the
curved line connecting the diamonds, while in the static case the
ordered region extends only down to $\rho\simeq 1$.

\section{Discussion}

A model (such as SSP) based on particles whose motion is biased by
fluctuations is likely to have a behaviour strongly dependent on
dimensionality around 2 dimensions since the critical dimension for
random walks is 2. An other facet of this aspect of the problem is that
a diffusing particle returns to the vicinity of any point of its
trajectory with probability 1, while the probability of for the same to
occur in 3d is less than 1. In other words, the diffusing particles
and clusters of particles are likely to frequently interact in 2d, but
in a three dimensional simulation they may not interact frequently
enough to ensure ordering.

Our calculations, however, show that for any finite density for small
enough noise there is an ordering in the SSP model.

On the other hand, in the 3d case it is very difficult to estimate the
precise value of the exponent describing the ordering as a function of
the noise.  The value we get within the errors agrees with the exponent
which is obtained for the equilibrium systems in the mean filed limit.
It is possible that the correlations in the direction of motion of the
particles spread so efficiently due to their motion that the SSP model
behaves already in 3d similarly to an infinite-dimensional static
system.  Indeed, the motion leads to an effective long-range
interaction, since particles moving in opposite direction will soon
get close enough to interact.

Finally, these findings indicate that the three dimensional SPP system
can be described using the framework of classical critical phenomena,
but shows surprising new features when compared to the analogous
equilibrium systems.  The velocity $v_0$ provides a control parameter
which switches between the SPP behavior ($v_0>0$) and equilibrium type
models ($v_0=0$).

\section*{Acknowledgments}

This work was supported by OTKA F019299 and FKFP 0203/1997.

\bfg
\caption{
The order parameter $\phi$ vs the noise amplitude
$\eta$ for the 3D SPP model for various system sizes. In these
simulations the density was fixed and the system size (numner of
particles $N$) was increased to demonstrate that for any system size the
ordered phase disappears in a continuous manner beyond a size dependent
critical noise.}
\label{Fig1} \efg

\bfg
\caption{(a) The order parameter $\phi$ vs the noise amplitude
$\eta$  (N=1000). (b)
As a comparison, when $v_0=0$ the behavior of the model is similar to
diluted ferromagnets: $\phi$ vanishes below the percolation
threshold ($\rho^*\simeq 1$).}
\label{Fig2} \efg

\bfg
\caption{
The diamonds show our estimates for the critical noise for a given
density of the particles in the SPP model and the crosses show the
critical noise for the static
case as a function of density. The SPP system becomes ordered in the
whole region below
the curved line connecting the diamonds, while in the static case the
ordered region extends only down to $\rho\simeq 1$ corresponding to the
percolation transition in the units we are using.
 }
\label{Fig3}
\efg

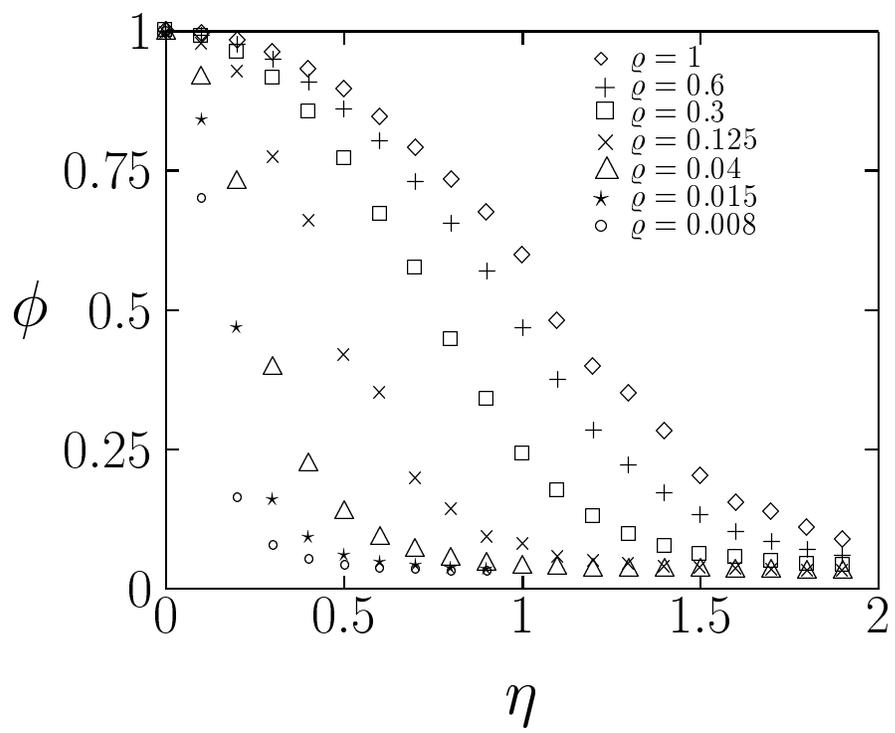
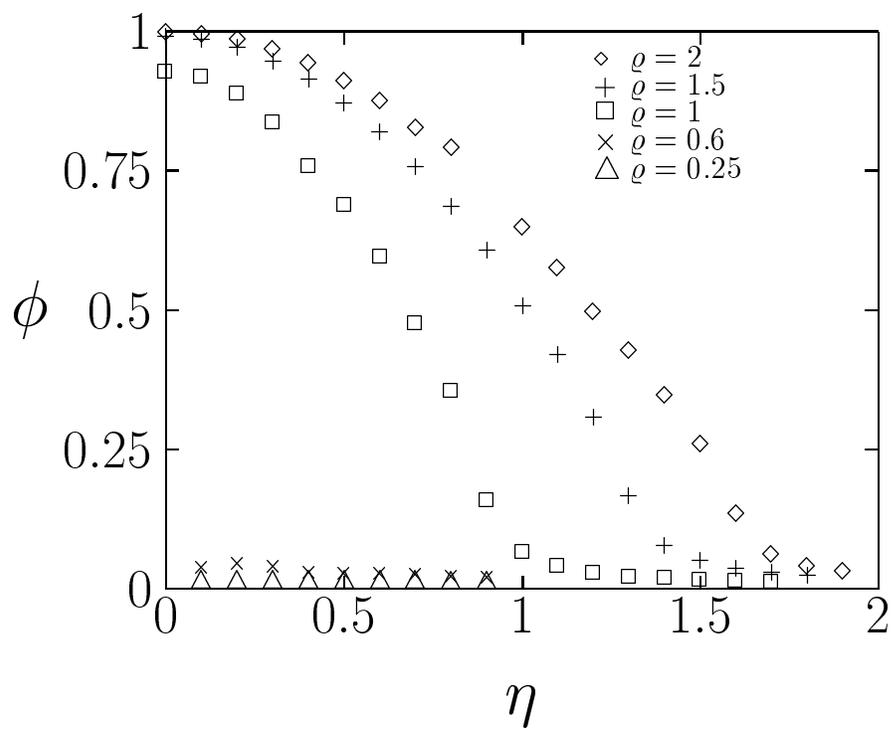
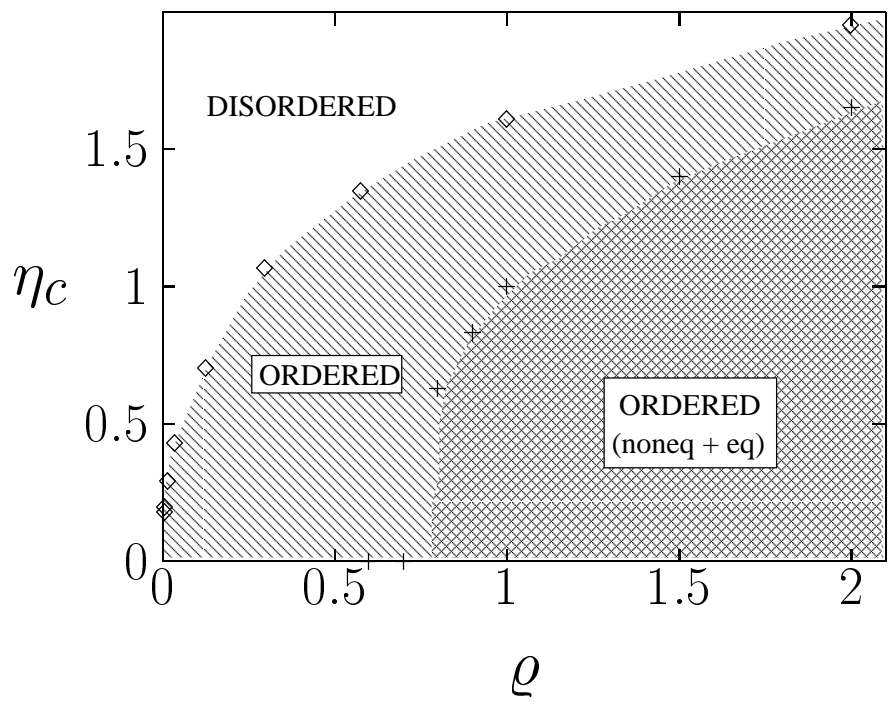
\end{document}